# Angle-resolved photoemission spectroscopy study of the charge density wave order in layered semiconductor EuTe$_4$

Chen Zhang,[1] Qi-Yi Wu,[1] Ya-Hua Yuan,[1] Wei Xia,[2,3] Hao Liu,[1] Zi-Teng Liu,[1] Hong-Yi Zhang,[1] Jiao-Jiao Song,[1] Yin-Zou Zhao,[1] Fan-Ying Wu,[1] Shu-Yu Liu,[1] Bo Chen,[1] Xue-Qing Ye,[1] Sheng-Tao Cui,[4] Zhe Sun,[4] Xiao-Fang Tang,[5] Jun He,[1] Hai-Yun Liu,[6] Yu-Xia Duan,[1] Yan-Feng Guo,[2,3,*] and Jian-Qiao Meng[1,†]

[1]*School of Physics and Electronics, Central South University, Changsha 410083, Hunan, China*
[2]*School of Physical Science and Technology, ShanghaiTech University, Shanghai 201210, China*
[3]*ShanghaiTech Laboratory for Topological Physics, ShanghaiTech University, Shanghai 201210, China*
[4]*National Synchrotron Radiation Laboratory, University of Science and Technology of China, Hefei 230029, Anhui, China*
[5]*Department of Physics and Electronic Science, Hunan University of Science and Technology, Xiangtan 411201, Hunan, China*
[6]*Beijing Academy of Quantum Information Sciences, Beijing 100085, China*
(Dated: Wednesday 16th March, 2022)

Layered tellurides have been extensively studied as a platform for investigating the Fermi surface (FS) nesting-driven charge density wave (CDW) states. EuTe$_4$, one of quasi-two-dimensional (quasi-2D) binary rare-earth tetratellurides CDW compounds, with unconventional hysteretic transition, is currently receiving much attention. Here, the CDW modulation vector, momentum and temperature dependence of CDW gaps in EuTe$_4$ are investigated using angle-resolved photoemission spectroscopy. Our results reveal that (i) a FS nesting vector $q \approx 0.67\ b^*$ drives the formation of CDW state, (ii) a large anisotropic CDW gap is fully open in the whole FS, and maintains a considerable size even at 300 K, leading to appearance of semiconductor properties, (iii) an abnormal non-monotonic increase of CDW gap in magnitude as a function of temperature, (iv) an extra, larger gap opens at a higher binding energy due to the interaction between the different orbits of the main bands.

The interactions among charge, spin, orbit and lattice in solids lead to abundant exotic quantum phenomena [1], such as CDW [2–4], high temperature superconductivity [5], heavy Fermions [6], and so on. These orders are often entangled with each other. For example, in the past two decades, the CDW order has been proven to have a complex interplay, competition or coexistence, with the superconducting order at low temperatures, especially in unconventional superconductors [7–11]. Understanding the formation mechanism of CDW is of great importance to understand their interaction with other orders, and is also the key to understand and control the collective motion of CDW orders.

In a one-dimensional system, the FS nesting mechanism proposed by Peierls has successfully explained the formation of CDW. Briefly, the FS nesting mechanism means that one section of FS can completely coincide with another FS after translating by a wave vector $q_{CDW}$, resulting in the redistribution of electron density in the new periodic field and the opening of a complete gap at Fermi energy ($E_F$) [3, 4, 12]. However, almost no perfect FS nesting occurs in two-dimensional (2D) and three-dimensional (3D) systems. The imperfect FS nesting causes partial energy gap opening and FS reconstruction, leaving residual electronic pockets so that the system maintains a metallic state in the CDW phase [13–16].

Telluride is the representative of chalcogenide compounds with a large number of family members, including monotellurides [17], ditelluride [18–21], tritellurides [14–16, 22–25], tetratellurides [26], etc. CDW states driven by FS nesting have been widely found in a large number of quasi-2D polytellurides [14–16, 21–24, 26], which makes it an ideal platform for studying CDW sequences.

Recently, a novel quasi-2D binary rare-earth tetratellurides with CDW phase, EuTe$_4$, was reported [27]. The CDW phase in EuTe$_4$ has been confirmed by transmission electron microscopy (TEM) and X-ray diffraction, which identified a in commensurate modulation wave vector $q = 0.643(3)\ b^*$ (where $b^* = \frac{2\pi}{b}$) [28]. Figure 1(a) shows the crystal structure of EuTe$_4$. In contrast to $R$Te$_3$ ($R$ = Dy - Lu) [14–16], an additional Te-Te square-net plane is inserted into the adjacent Eu-Te slabs, which causes EuTe$_4$ to contain two types of Te planes: bilayer and monolayer separated by the EuTe layer. It was considered that the CDW modulation of rare-earth tellurides comes from Te square layers [15, 16]. Thus, there would be a relative difference in CDW distortion between these two different types of Te square layers, which causes some novel characteristics associated with CDW that are significantly distinct from the $R$Te$_3$ compounds [28, 29]. As reported, a large temperature hysteresis loop was observed in EuTe$_4$ [27, 28], which suggestes a close relation to 3D configurations of the in-plane density waves [28]. The temperature-dependent resistivity measurements on EuTe$_4$ show typical semiconductor behavior rather than metallic behavior in the CDW state [27, 28]. Theoretical calculation of electronic susceptibility suggests that the monolayer and bilayer Te layers primarily contribute to FS nesting and electronic instability, respectively, to drive the formation of CDW in EuTe$_4$ [29].

In this letter, we present a detailed angle-resolved photoemission spectroscopy (ARPES) study of the electronic structure and CDW gap in EuTe$_4$ as a function of temperature and momentum. It is found that a large anisotropic CDW gap is fully open in the whole FS and therefore had no spectral weight around $E_F$, thus resulting in its semiconducting properties. The low-lying electronic structure is the linearly dispersive bands that are dominated by the Te 5$p$ band derived from Te square nets. We consider that the imperfect FS nesting drives the formation of CDW state in EuTe$_4$. The data show that the CDW gap has a very strong momentum and temperature dependence. Moreover, the CDW gap persists even at room temperature, and its temperature dependence is non-monotonic, which obviously deviates from the prediction of traditional CDW theory.

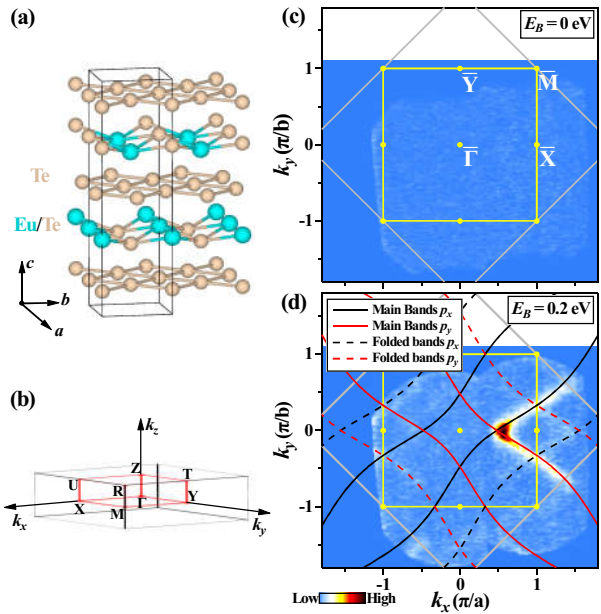

FIG. 1. (**a**) The crystal structure of EuTe$_4$. (**b**) A 3D reduced Brillouin Zone (BZ) of EuTe$_4$ with high symmetry points marked. (**c**) and (**d**) Constant energy contours at the $E_F$ and at 0.2 eV below $E_F$ as a function of $k_x$ and $k_y$ measured at a temperature of 30 K. The solid yellow squares outline the projected first BZ of the 3D crystal structure, and the solid grey squares represent the extended BZ formed by Te square net. The black ($p_x$) and red ($p_y$) solid lines in (d) are the expected bands from the Te square-net calculated by the TB model, and the corresponding dashed lines represent the folded bands.

High-quality EuTe$_4$ single crystals were grown via the Te self-flux approach. ARPES measurements were performed at the ARPES beamline (BL13U) of the National Synchrotron Radiation Laboratory, Hefei, with a Scienta Omicron DA30L analyzer. All samples used in this study were cleaved *in situ* and measured in an ultrahigh vacuum with a base pressure better than $6 \times 10^{-11}$ Torr. The energy and angular resolutions were better than 20 meV and 0.3°, respectively. The Fermi level is referenced by measuring clean polycrystalline Au that is electrically connected to the sample. All data were taken with 35 eV photons.

As presented in Fig. 1(c), the measured FS of EuTe$_4$ at 30 K shows no intensity contrast, indicating that the whole FS is fully gapped. This corroborates that it exhibits semiconductor characteristics in transport measurements. Figure 1(d) displays the constant energy contour at binding energy $E_B = 0.2$ eV, and an apparent twofold symmetry of the low-lying electronic structure can be seen. Calculation suggested that the $4f$ electrons of Eu are localized around 1.5 eV below $E_F$ [27], and the electronic states near the $E_F$ are dominated by the $5p$ orbitals of Te atoms on the Te-Te plane [27–31]. Thus, similar to the $R$Te$_3$, the low-lying band structure of EuTe$_4$ can be described by an elementary 2D tight-binding (TB) model, which only includes the in-plane $p_x$ and $p_y$ orbitals of the Te plane and contains no hybridization between them [15, 16, 32]. Using the axes of the 3D BZ, this model yields dispersions for $p_x$ and $p_y$ as the following equations [16, 23]:

$$E_{p_x}(\mathbf{k}) = -2t_\parallel \cos\left[(k_x + k_y)\frac{a}{2}\right] + 2t_\perp \cos\left[(k_x - k_y)\frac{a}{2}\right] - E_F,$$

$$E_{p_y}(\mathbf{k}) = -2t_\parallel \cos\left[(k_x - k_y)\frac{a}{2}\right] + 2t_\perp \cos\left[(k_x + k_y)\frac{a}{2}\right] - E_F.$$

where the $t_\parallel$ and $t_\perp$ are the hopping amplitude along and perpendicular to the $p_x$ orbital, respectively. Otherwise, considering that this TB model is constructed on a single Te plane which is rotated 45° with respect to the unit cell and has only half the area. Therefore, the area of the 2D BZ [grey squares in Fig. 1(d)] for the TB model is twice of the 3D BZ [yellow squares in Fig. 1(d)], and its periodicity should be folded back with respect to the 3D BZ boundaries to acquire the 3D lattice symmetry. This gives rise to the folded bands, which are along the reduced BZ boundaries are shown by the dashed lines in Fig. 1(d). Figure 1(d) compares the measured and simulated FS cross sections at $E_B = 0.2$ eV. It can be seen that the measured and simulated FS cross sections agree to each other very well. On the 3D BZ boundary, a $\bar{X}$-centered diamond-like pocket formed by crossing main bands and its folded bands is clearly observed, similar as EuSbTe$_3$ [33].

Next, we study how the FS contours change for energies below the $E_F$; the resulting maps are given in Figs. 2(a1)-2(a4). At each $E_B$, the TB model and experimental data are excellently consistent, suggesting this model is reasonable for understanding the band structure of EuTe$_4$. As $E_B$ increases, the $\bar{X}$-centered diamond-like pocket keeps shrinking. On the contrary, the spectral intensity around the $\bar{\Gamma}$ point gradually increases, and a larger inner diamond-shaped pocket is gradually formed. The band structure at high $E_B$ is reminiscent of the FS of $R$Te$_3$ [16].

Besides the main and folded bands, there are some weak features that we refer to as shadow bands, as indicated by the arrows in Fig. 2 (a1). Figure 2(b) shows a series of momentum distribution curves (MDCs) at $E_B = 0.4$ eV along seven evenly spaced cuts (cuts 1 to 7) as labelled in Fig. 2(a2). It can be seen that the distance between the shadow band peak and main band peak has little variation, that is, the shadow band is nearly parallel to the upper branch of the main band. The shadow band can be roughly produced by shifting the main band by a CDW wave vector $\boldsymbol{q}_{CDW} \approx \frac{2}{3}\boldsymbol{b}^*$, which is basically consistent with the TEM results [28].

To understand the implications of the transverse CDW on the band structure, a cut [cyan arrow in Fig. 2(a1)] parallel to the $k_y$ direction at $k_x = 0.35$ $\boldsymbol{a}^*$ ($\boldsymbol{a}^* = \frac{2\pi}{a}$) is taken. The obtained image and the corresponding energy distribution curves (EDCs) are presented in Figs. 2(c) and 2(d), respectively. Combined with the TB model, we can clearly distinguish the dispersion of the main bands (thick solid lines), the folded bands (dashed lines), and the shadow bands (thin solid lines). Two gaps, $\Delta_1$ and $\Delta_2$, can be observed, as indicated by black double-headed arrows in Fig. 2(d). $\Delta_1$ is opened near higher binding energy of 0.7 eV. Similar gaps have been observed in some layered rare-earth tellurides, which occur at the crossing point between the main bands and the folded bands [15, 16, 33]. However, in EuTe$_4$, this gap is not a result of the intersection between the main bands and the folded bands but the intersection between the different orbits of main bands. $\Delta_2$ is located around $E_F$, which is generally called the CDW gap. As previously reported in many FS nesting-driven CDW materials, the interaction between the main bands and the shadow bands leads to the opening of the CDW gap $\Delta_2$ and leaves a lower branch below $E_F$. This phenomenon further provides evidence of the validity of FS nesting in determining the CDW instabilities and the semiconducting property in this system.

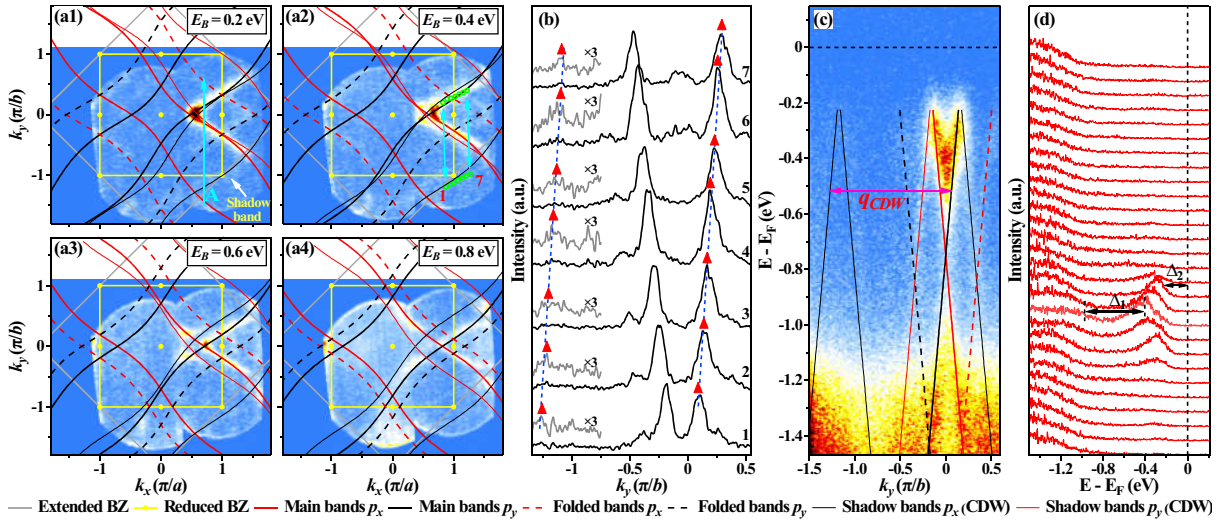

FIG. 2. (**a1-a4**) The constant energy maps integrated over 20 meV, centered at binding energies $E_B = 0.2, 0.4, 0.6,$ and 0.8 eV, respectively. (**b**) MDCs spectra at $E_B = 0.4$ eV for evenly spaced cuts (1-7) as shown in (a2). Note that the signal of the shadow bands is very weak. To see the weak features more clearly, the original MDCs (black lines) are expanded three times (grey lines). The curves are shifted vertically for easy view. The blue dashed lines serving as guide for the eyes. (**c**) ARPES image plot of cut A indicated by the cyan arrow shown in (a1). The straight solid and dashed lines represent band dispersion serving as guides for the eyes (see legend). (**d**) EDCs corresponding to (c), from where two gaps can be observed.

To quantitatively study the momentum dependence of the gaps, Figs. 3(a1)-3(a4) display the band structure along four cuts (cut 1 to 4) as labelled in Fig. 3(b). Both $\Delta_1$ and $\Delta_2$ vary significantly with momentum. The closer to the apex of the main bands (point A), the larger the $\Delta_1$ is. As shown in Fig. 2(d), the gap size of $\Delta_1$ can reach $\sim 600$ meV. When approaching the 3D BZ boundary, the intersection between the main bands becomes deeper and deeper and mixes with the $4f$-electron state of Eu, resulting in $\Delta_1$ that is almost closed or indistinguishable. However, the CDW gap $\Delta_2$ seems to present a completely opposite trend. Spectra along the main band formed 'underlying FS' are shown in Fig. 3(c) together with spectra (dotted line) from a polycrystalline Au foil in electronic contact with sample. The size of gap $\Delta_2$ is extracted from the leading-edge midpoint position of EDCs, indicated by a black double-headed arrow. $\Delta_2$ shows an obviously momentum dependent. Figure 3(d) summarizes the gap size as a function of $k_y$, which shows that the gap value is the largest near the 3D BZ boundary and decreases drastically while approaching the apex of the main band. At 30 K, the maximum and minimum gap sizes are $\sim 180$ meV and $\sim 130$ meV, re-spectively. At 240 K, the maximum gap value can reach $\sim 230$ meV, and the minimum gap value is $\sim 170$ meV. The variation of the gap along the 'underlying FS' was always observed in the FS nesting-driven CDW systems, which commonly arises from imperfect nesting [16, 23]. Furthermore, another factor needs to be considered, that is, the opening of $\Delta_1$ would inevitably push parts of spectral weight upwards, thus narrow-ing the distance between the EDCs leading-edge and $E_F$, especially around the apex of the main band. Besides, we also note an unusual phenomenon that the gap size does not evolve monotonously with temperature.

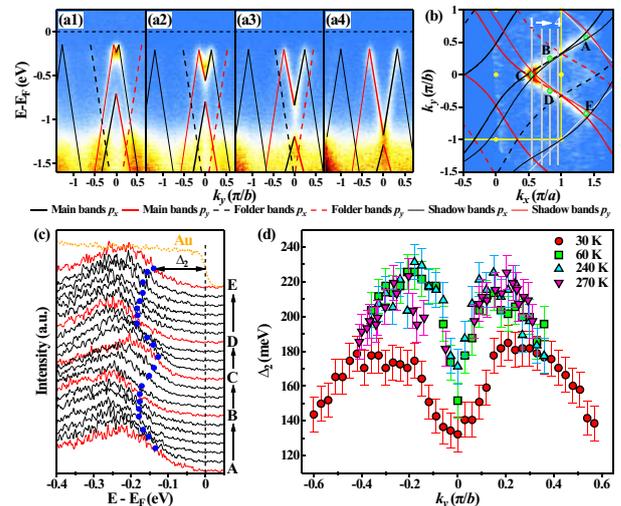

FIG. 3. (**a1-a4**) ARPES image plots along the $k_y$ direction for $k_x = 0.55, 0.7, 0.82,$ and 0.95 (in $\pi/a$ units) at $T = 60$ K. The lines represent band dispersion serving as a guide for the eyes (see legend). (**b**) Location of the cuts of (a) present in the constant energy contour at $E_B = 0.2$ eV. (**c**) EDCs at the main band from A to E at 30 K, along with reference Au spectra (doted line). The points A, B, C, D, E are represented by empty circles as shown in (b). (**d**) Momentum-dependent energy gap $\Delta_2$ at 30 K, 60 K, 240 K, and 270 K. The gap size is defined with respect to the leading edge of the spectra.

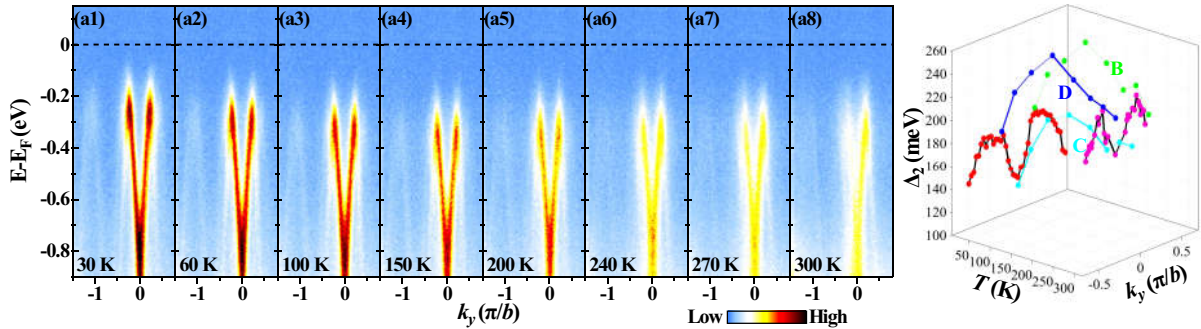

FIG. 4. (**a1-a8**) ARPES data of EuTe$_4$ along the same direction as in Fig. 3(b) at the labeled temperatures. (**b**) Summary of the CDW gap $\Delta_2$. Green, cyan, and blue dots represent the temperature-dependent CDW gap at points B, C, and D, as indicated in Fig. 3(b), respectively. The red dots show the momentum-dependent CDW gap at 30 K and 270 K.

Finally, the temperature evolution of the CDW gap was studied. Figures 4(a1)-4(a8) show detailed temperature evolution of the ARPES images measured along a representative momentum cut as Fig. 3(a3) from 30 K to 300 K. $\Delta_2$ remains open at 300 K, which is consistent with the CDW phase transition occurs well above room temperature in EuTe$_4$ [27, 28]. A remarkable feature is that $\Delta_2$ changes non-monotonically with temperature. Figure 4(b) summarizes the extracted $\Delta_2$ gap sizes at point B (green) and D (blue), as well as at point C (cyan). As temperature increases, the gap size keeps increasing until ∼ 150 K, and then gradually decreases. Such behavior obviously deviates from the prediction of traditional CDW theory, which suggests the gap should follow a BCS curve [4]. A similar abnormal temperature-dependent CDW gap was also observed in EuSbTe$_3$, and the temperature evolution of the higher binding energy gap was considered to plays a role [33]. However, this scenario does not apply here. First, the higher binding energy gap $\Delta_1$ is almost closed at all temperatures at the momentum we measured, which could not affect the CDW gap $\Delta_2$. Second, the CDW gaps at measured temperatures have similar momentum dependence [Fig. 3(d) and Fig. 4(b)], which means that the temperature evolution of the CDW gap at the 'underlying FS' should be the same. This further proves that the anomalous temperature dependence of CDW gap $\Delta_2$ is independent of the high energy gap $\Delta_1$. Furthermore, as mentioned above, a relative difference exists in the phase and mechanism in CDW distortion between the Te square bilayer and monolayer [28, 29]. Although this effect is not distinguished from the evolution of electronic structure. But, as previously reported in EuTe$_4$, the interaction between the bilayer and the monolayer CDWs may lead to hysteresis and a relatively large CDW gap in the heating branch [28]. However, it is worth noting that the CDW gap has begun to increase with the increase of temperature before the onset temperature of hysteresis. Therefore, there may be other reasons that lead to the sharp decrease of CDW gap at low temperatures, such as the emergence of a hidden order which suppresses the CDW. More experimental and theoreti-cal works are required to explore such an abnormal behavior of the temperature-dependent CDW gap.

To conclude, based on the systematic ARPES measurement, a detailed analysis of the observed electronic structure on EuTe$_4$ single crystals was presented, suggesting that the FS nesting along $b^*$ drives the CDW modulation results in its FS reconstruction and leaves a relatively large CDW gap that opens over the whole FS. The CDW modulation vector $q \approx 0.67 b^*$ is exact from a two-fold symmetric low-lying electronic structure. At higher binding energy, the interaction between the different orbits of the main bands introduces an extra band gap, $\Delta_1$, which should be one of the main reasons for the anisotropic CDW gap. Moreover, the temperature-dependent data reveal that the CDW gap exhibits abnormal behavior at all momenta with temperature increases. Our findings provide detailed information and critical insight into understanding of FS nesting-driven CDW states on the electronic structure.

This work was supported by the National Natural Science Foundation of China (Grant No. 12074436 and No. 11874264), the Innovation-driven Plan in Central South University (Grant No. 2016CXS032), and the Natural Science Foundation of Shanghai (No. 17ZR1443300).

---

* Corresponding author: guoyf@shanghaitech.edu.cn
† Corresponding author: jqmeng@csu.edu.cn